\documentclass[amsmath,amssymb,lengthcheck,singlecolumn,aps,prl] {revtex4}
\usepackage[T1]{fontenc}
\usepackage[utf8]{inputenc}
\usepackage{graphics}
\usepackage{graphicx}

\usepackage{xcolor}

\begin{document}
\title{\textbf{Collective excitations of Bose-Einstein Condensate in a Rydberg Atom}}
\author{Avra Banerjee}

 \author{Dwipesh Majumder}
\affiliation{Department of Physics, Indian Institute of Engineering Science and Technology, Shibpur, W B, India}

\begin{abstract}

In this work, we investigated a  Bose-Einstein condensate (BEC) inside  a Rydberg atom.
We first observed the system's density profile using the Gross-Pitaevskii equation, and then we used the Bogoliubov theory to examine the collective excitation spectra.
We have also extended our study by taking into account a hybrid of two BEC species.
Again, for this new system, the density profile and the collective excitation are investigated in a similar manner.
The results of the BEC for one species reveal that the Rydberg atom can trap the BEC and that the excitation curve moves upward with increasing Rydberg atom interaction.
We have seen roton minima for two species of BEC, and the depth of the minima shifts with the Rydberg  interaction. 
\end{abstract}

\maketitle

\section*{Introduction}

Bose-Einstein condensation (BEC) of ultracold atoms confined in a Rydberg atom\cite{Polaron_The2016, Polaron_Exp2018}, known as Rydberg polaron is a very interesting phenomenon, where an electron confines thousands of atoms. Usually, we discuss an electron of an atom, but here the scenario is completely different. The first Rydberg atom in ultracold atomic BEC was proposed (theory)  by Greene et al. in 2000\cite{PRL85.2458}.
There is a possibility of application of an array of Rydberg atoms in quantum computation\cite{QComputer}, which gives additional importance to studying the Rydberg atoms in the ultracold atomic system.
Nowadays, researchers routinely create giant Rydberg atoms with principal
quantum number around 200\cite{ryd_pqn}, which corresponds to an orbital radius of
several micrometres, which can confine thousands of Bose atoms at the ultra-cold temperature. 

Many aspects of the Rydberg atom in BEC have been investigated so far. Here, we are listing some of them for completeness.
There are some studies looking at the existence of impurities (ion or Rydberg atom) in the BECs at ultracold temperatures \cite{imp_bec1,imp_bec2}. Advanced techniques for controlling ion-atom combinations, such as cold collision, chemical reactions \cite{cold_chem1,cold_chem2,cold_chem3,cold_chem4}, and single ions in BEC \cite{single_ion}, have been studied.
The motion of Rydberg atom has been studied by tracking the Rydberg atom in the BEC using phase variation due to the motion of the Rydberg atom.
  With the aid of the related phase data, we may understand the Rydberg atom's motion in the mixture \cite{ryd_dy6}. The impurity-based BEC can be used as a platform for the research of micro-macro, and macro-micro entanglement  \cite{entan1,entan7}. Furthermore, BEC perfectly controls the micro-macro quantum system  \cite{mic_mac}. A system of ultracold Rydberg atoms submerged in atomic BEC was studied by Jia Wang et al. \cite{wang}.They have demonstrated that BEC can image the Rydberg electron for modest healing lengths while being appealing for high healing lengths.
Rydberg atoms are bound together by the Yukawa interaction. By observing how the Rydberg electron interacted with the condensate, SK Tiwari et al. were able to determine the position and speed of moving Rydberg atoms \cite{wuster}.
They explained how to examine ionization collision using this tracking technique. Theoretically, Rammohan et al. explained the decoherence caused by phonons in the superposition of two separate Rydberg states \cite{rammohan}.
The Rydberg phonon coupling coefficient has been discovered. They determined the scale of
the bath correlation function.

Liquid formation of the dilute ultracold atomic system \cite{Petrov2015,Trarruell2018}  is another exciting topic in Bose-Einstein condensation (BEC).
The droplets have been observed in the isotropic short-range interacting system of two species of cold atoms \cite{Trarruell2018,drop_exp2, Trarruell2018PRL} as well as in the anisotropic long-range dipolar interacting system of $^{164}$Dy or $^{166}$Er atoms \cite{dipolar_droplets}. 
In the mixture of two-component Bose atoms, the spherical droplet has been observed under the competition between the effective short-range attractive interaction and the repulsive interaction due to the quantum fluctuation \cite{LHY}.
The two-component BEC may be the mixture of atoms of two different elements (different atomic mass) \cite{PRL89, PRL100,Itali2020} or maybe the mixture of atoms with two different internal degrees of freedom of a given element \cite{spin_BEC, PRL'101}. 

Here, we shall discuss the collective excitation of BEC in the gas phase as well as in the liquid phase confined within the Rydberg atom. In our study, we have considered single-species BECs (in the gas phase) as well as a mixture of two species BECs of atoms of the same isotope but different internal degrees of freedom (in the gas phase and the liquid phase). Here, we represent the collective excitation of the system. 

The fundamental structure of a BEC could be well approximated by mean-field theory, namely the Gross-Pitaevskii (GP) energy functional, due to the low atomic energies and densities attained in the experiments.
In these circumstances, the effects of s-wave collisions dominate the atom-atom interaction. While we have, in this instance, ignored the harmonic trapping potential, the Rydberg atom still allows us to trap the BEC up to a limit. Then, we explored what happens when the BEC  is subjected to perturbation and studied the collective excitation using the Bogoliubov approach. 
It is simple to study the elementary excitation in the uniform BEC of infinite size using Bogoliubov theory.
The well-known phonon mode is the low energy collective excitation in this weakly interacting system of an ultracold dilute gas of Bose atoms in the condensate. 
We extended our study to include the BEC quantum droplets in the Rydberg atom. In general, the ultracold, diluted Bose condensate shows phonon modes in the excitation spectra in the low-energy excitation regime. The quantum droplet of BEC has completely different properties than normal BEC. The size of the droplet is quite large, and the density is constant over a long region except at the surface. We have done our studies of collective excitations in three dimensions with this Rydberg dressing for both single and double species.


\section*{ I. Model and calculation}

\subsection{A. BEC confined in a Rydberg atom}

 Usually, a system of a finite number of repulsive Bose atoms cannot be kept together without any trapping potential. We have taken into account the Rydberg atom's electron contribution in our work because the ion-atom polarization potential changes inversely as the fourth power of radial distances \cite{qd5}. We consider the repulsive Rydberg electron-atom interaction to confine the atoms. In the presence of the Rydberg atom, the density of the Bose gas at absolute zero temperature is almost constant due to the nature of the Rydberg-electron cloud distribution. Here we are considering one species of Bose gas that forms Bose-Einstein condensation, which follows the mean field GP-equation\cite{GP_Ry1, GP_Ry2}.

\begin{equation}
i\frac{\partial \psi}{\partial t}=\left [-\frac{ \nabla^2}{2}+g |\psi |^2+g_{LHY}|\psi |^3+V_0|\Psi (r)|^{2}\right ] \psi
\end{equation}
here $\psi$ is the order parameter wave function of the condensate.
On the right, the first term is the kinetic energy term, the second is the interatomic interaction term with interaction strength $g$ (weak repulsive contact interaction has been considered, and the third is the correction over mean-field interaction potential ($g_{LHY} = \frac{32g}{3} \sqrt{\frac{a^3}{\pi }}$ \cite{qd3} ; $a$ is the scattering length of the interaction,$g=4\pi a$) term due to quantum fluctuations and the last term ($V_0 = \frac{2\pi m a_e}{m_e }$ \cite{wang,wuster} and $a_e$ is electron atom scattering length)for  Rydberg potential contribution with electron-atom interaction strength $V_0$ ; $\Psi(r) = \Psi_{n00}(r)$ is the hydrogen atom wave function with large principle quantum number.
Here we have expressed the length in unit of $l_{0}=0.1 \mu m$ ,  time in $\frac{ml_0^2}{\hbar}$, energy in $\frac{\hbar^{2}}{ml_{0}^{2}}$ unit , $m$ is considered as the mass of the condensed atom, $m_e$ mass of an electron.As our system is radially symmetric in spherical coordinates, we choose $\nabla^2 =\frac{1}{r^2} \frac{\partial}{\partial r}\left( r^2\frac{\partial}{\partial r} \right)$. It is a standard procedure to write the wave function as $\psi = \phi(r) /r$ . The GP-equation takes the form,
\begin{equation}
i\frac{\partial \phi}{\partial t}=\left [-\frac{1}{2}\frac{ \partial^2}{\partial r^{2}}+g |\frac{\phi}{r} |^2+g_{LHY}|\frac{\phi}{r} |^3+V_0 |\Psi (r)|^{2}\right ] \phi
\end{equation}
We have used the imaginary-time split-step Crank Nicolson method to solve the GP-equation  \cite{adhikary} with the boundary conditions $\phi(r=0)=0$ and $\phi(r=\infty)=0$ and  normalization
  \begin{eqnarray}
   4\pi  \int_0^\infty |\phi|^{2} dr= N
  \end{eqnarray} 
where N denotes the total number of particles present in the condensate.
In actual numerical calculation, we cannot consider the infinite distance; rather, we consider a sufficiently large distance as the boundary.

 The Rydberg atom is capable of confining the condensate. In this case, both the density and radial extension of the condensate increase if the number of particles increases. The ground state density is shown in FIG. 1.

\begin{figure} [htbb]
  \begin{center} 

  \includegraphics[width=0.50\textwidth]  {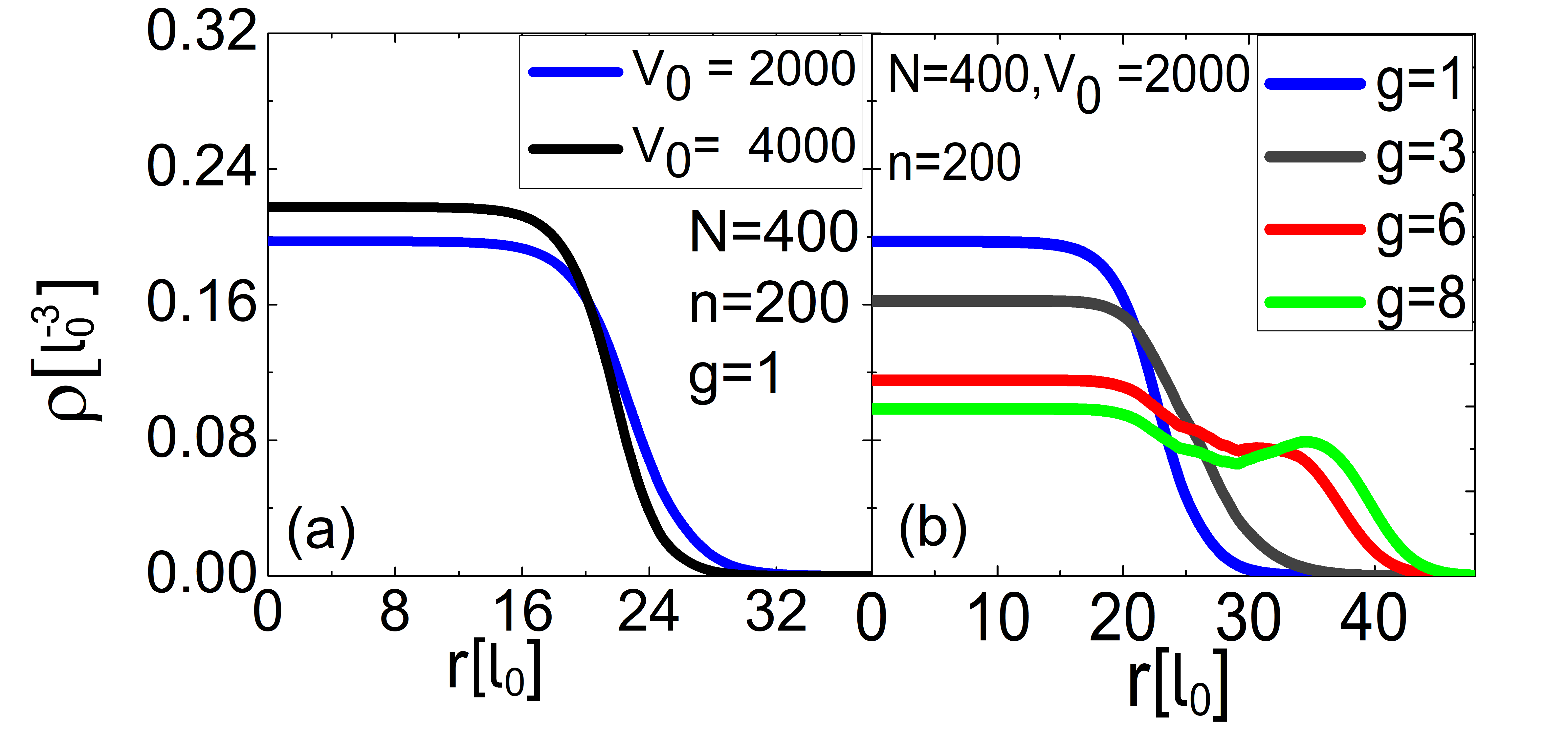}
  
\vspace{0.2 cm}
  \includegraphics[width=0.48\textwidth]  {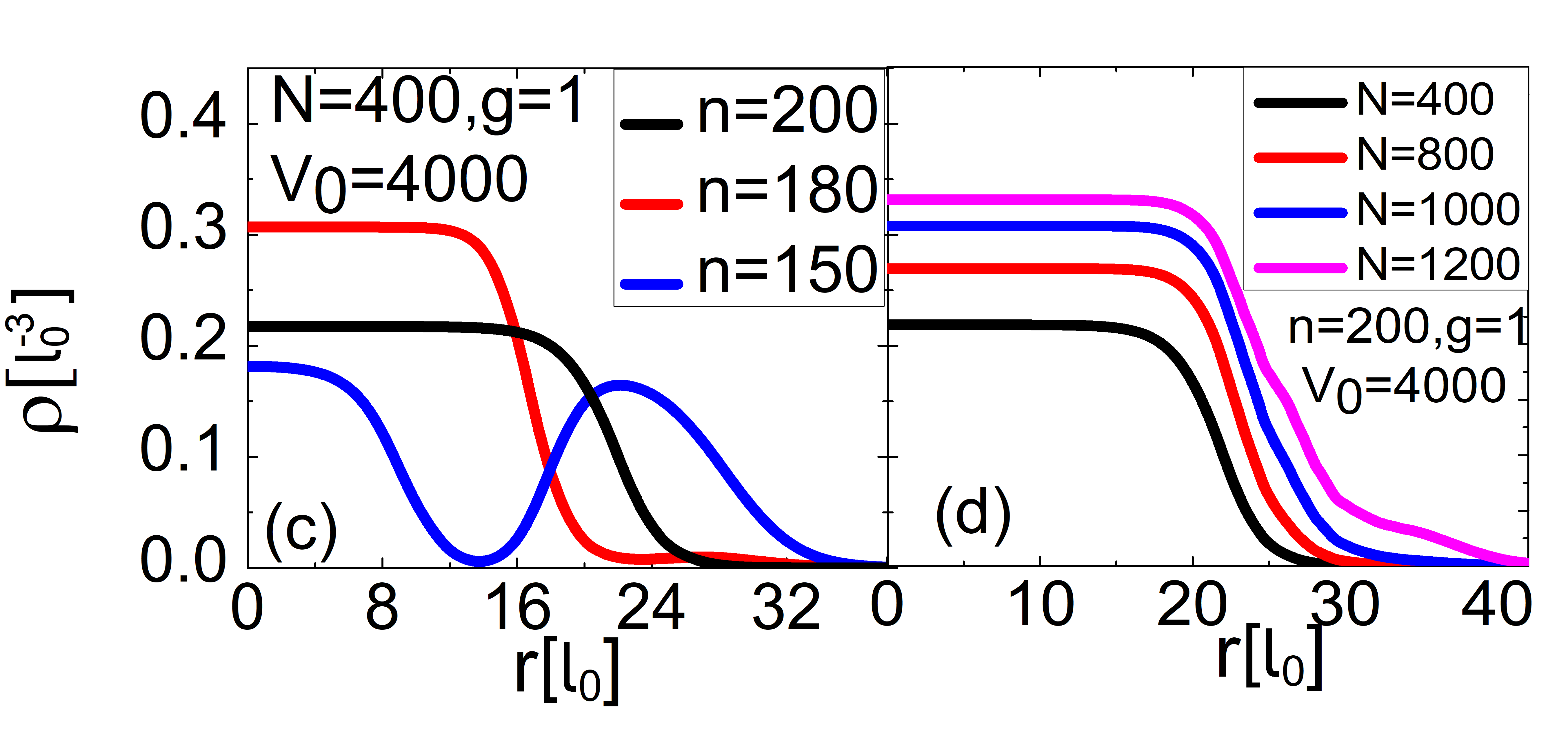}

\caption{Density of the single species BEC inside the Rydberg atom. \textbf{(a)}: If we increase $V_0$ the density of the condensate inside the Rydberg atom also increases; \textbf{(b)}: If we increase the interaction strength $(g) $, the pressure inside the Rydberg atom will be high, and some atoms will move out of the Rydberg atom;
 \textbf{(c)}: If we reduce the size of the Rydberg atom, some atoms of the condensate will move out of the Rydberg atom;
 \textbf{(d)}: There is a limit to confining the numbers of the atoms in the condensate.}
   \end{center}
   \label{fig:1den_V0}
\end{figure}

For the Rydberg atom, we chose a very large principal quantum number so that the Rydberg electron density is far from the central region. Effectively, the probability of finding a Rydberg electron near the central region is negligible.
\subsection{Collective excitation}

Our condensate is about 40$l_0$ in diameter, whereas the coherence length is about $l_0$, so the size of the condensate is sufficiently large to study the collective excitation of the system in the central region, where the density of the condensate is almost uniform.
So, we are using the Bogoliubov theory to find the collective excitation of the condensate . The collective excited state can be considered as
 $\psi^{\mbox{exc}} = \psi+\delta \psi$. The perturbation part can be written as \cite{dm_coll, ch_bose},

\begin{footnotesize}
\begin{equation}
  \delta \psi = e^{-i\mu t} \left (U e^{i\vec{q}\cdot  \vec{r} -i\omega t} + V^* e^{-i\vec{q}\cdot  \vec{r} +i\omega t}\right )
\end{equation}
\end{footnotesize}
\begin{small}
Where chemical potential is given by \cite{adhikary},
\begin{equation}
\mu =4\pi\int_{0}^\infty \biggl [ \frac{1}{2}|\frac{\partial   \phi}{\partial  r} |^{2}+\frac{g}{r^{2}}|\phi|^{4}+\frac{g_{LHY}}{r^{3}}|\phi|^{5}+ V_0 |\Psi (r)|^{2}|\phi|^{2}\biggl ]  dr
\end{equation}
\end{small}
 $\{U, V\}$ are amplitude of the excitation, $\vec{q}$ is quasi-momentum of the excitation. If we put these in our GP equations, we will find the equations for $\{U, V\}$, in the first order approximation of $\{\delta\psi\}$.

\begin{figure} 
  \begin{center} 
  \includegraphics[width=0.5\textwidth]  {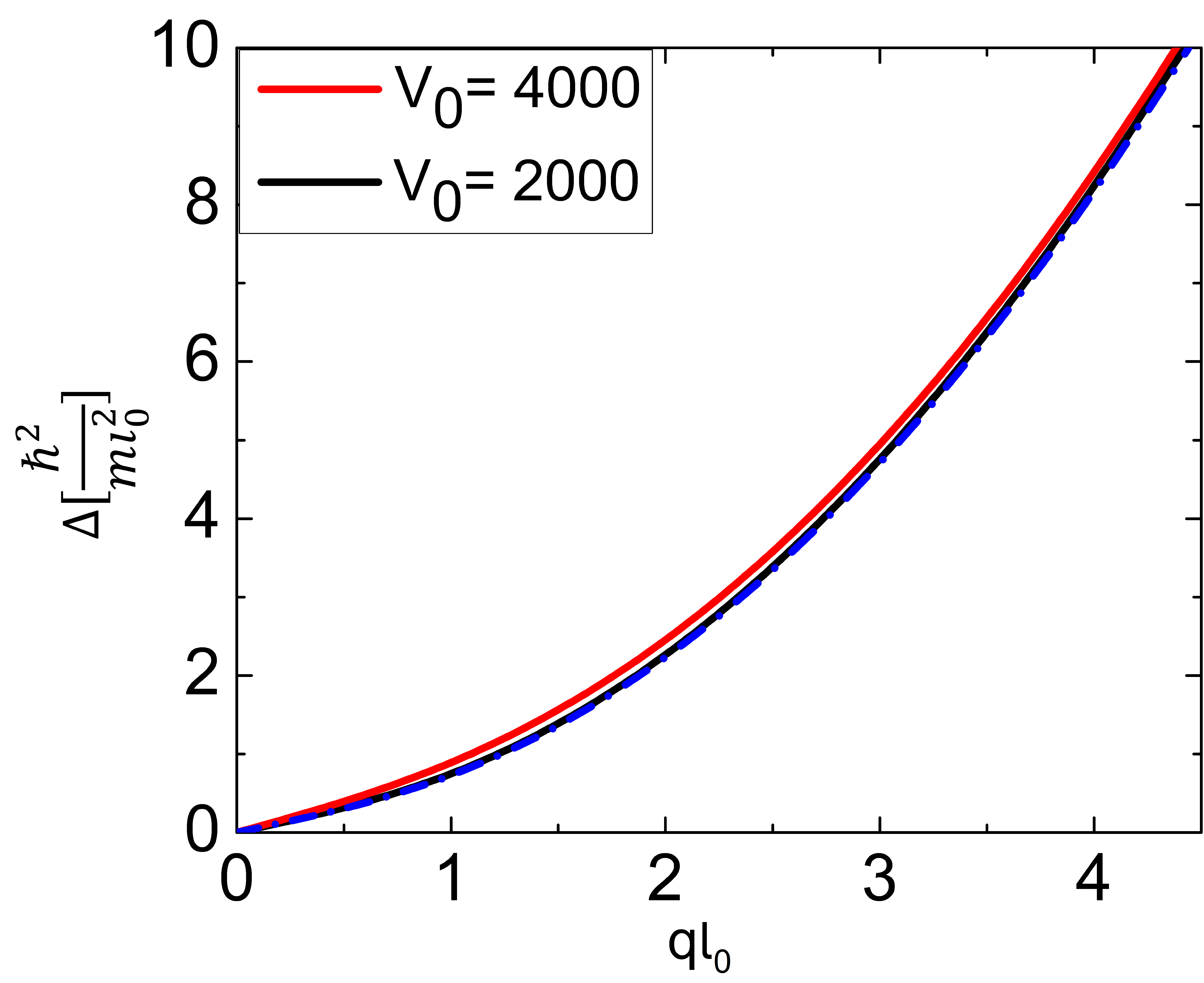}
\caption{Collective excitation spectra  of single species BEC with Rydberg atom for various values of $V_{0}$ and $g=1$. Dashed blue excitation curve for the uniform BEC  with the density as the black curve ($\rho=0.197$). }
   \end{center}
\end{figure}

\begin{footnotesize}
    \begin{eqnarray}
\left(
    \begin{tabular}{c c c c}
      $h_{1}$ & $g\psi^2+\frac{3}{2}g_{LHY}|\psi|^3$\\
       $-(g\psi^2+\frac{3}{2}g_{LHY}|\psi|^3)$ & $-h_{1}$  \\

    \end{tabular}
\right)
\left(
\begin{tabular}{c}
$U$ \\$V$
\end{tabular}
\right) = \omega \left(
\begin{tabular}{c}
$U$ \\$V$ 
\end{tabular}
\right) 
\end{eqnarray}
where
\begin{eqnarray}
 h_{1}= \frac{q^2} {2}+ 2g |\psi|^2+\frac{5}{2}g_{LHY}|\psi|^3-\mu+V_0 |\Psi (r)|^{2}
 \end{eqnarray}

  \end{footnotesize}
  
We obtained the collective excitation by diagonalizing equation (6).

\begin{small}
\begin{eqnarray}
 \Delta=\sqrt{ h_{1}^2-(g\psi^2+\frac{3}{2}g_{LHY}|\psi|^3)^2}
\end{eqnarray}
\end{small}

The excitation spectra have an almost parabolic nature, and for greater Rydberg scattering lengths, the spectra move to higher energy regions.


\section*{B. Mixture of BECs in a Rydberg Atom}

Now we have extended our study to a mixture of two species of BEC. We consider the mixture of two kinds of Bose atoms of the same isotope of an element to have two different internal degrees of freedom. 


 In the ultracold temperature and very low density the interaction potential between two nutral atoms can be written as $V(\vec{r}_1,\vec{r}_2) = g_{ij} \delta(\vec{r}_1-\vec{r}_2)$. 

With this natural unit, the coupled non-linear GP equations with the LHY interaction and Rydberg dressing in two species can be written as \cite{ref10}

\begin{footnotesize}
\begin{eqnarray}
    i\frac{\partial \psi_i}{\partial t}=[-\frac{\nabla  ^2}{2}+g_{ii}|\psi_i|^2-g_{12}|\psi_j|^2
  +g_{LHY}|\psi_i|^3 +V_0  |\Psi(r)  |^2]\psi_i
\end{eqnarray}
\end{footnotesize}

The first term on the right side is the kinetic energy term, the second term is intraparticle interaction, the third term is interparticle interaction, the fourth cubic term is responsible for quantum fluctuation, and the last term is the Rydberg atom contribution.

After the transformation $\psi_i = \phi_i(r) /r$, the GP equations reduce to

\begin{eqnarray}
    i\frac{\partial \phi_i}{\partial t}=\biggl [-\frac{1}{2}\frac{\partial  ^2}{\partial r^2}+g_{ii} \frac{|\phi_i|^2}{r^2}-g_{12} \frac{|\phi_j|^2}{r^2}  \\
  +g_{LHY}|\frac{\phi_i}{r}|^3 +V_0 |\Psi(r)  |^2 \biggr ]\phi_i \nonumber
\end{eqnarray}

     \begin{figure} 
   \begin{center}
     \includegraphics[width=0.49\textwidth]{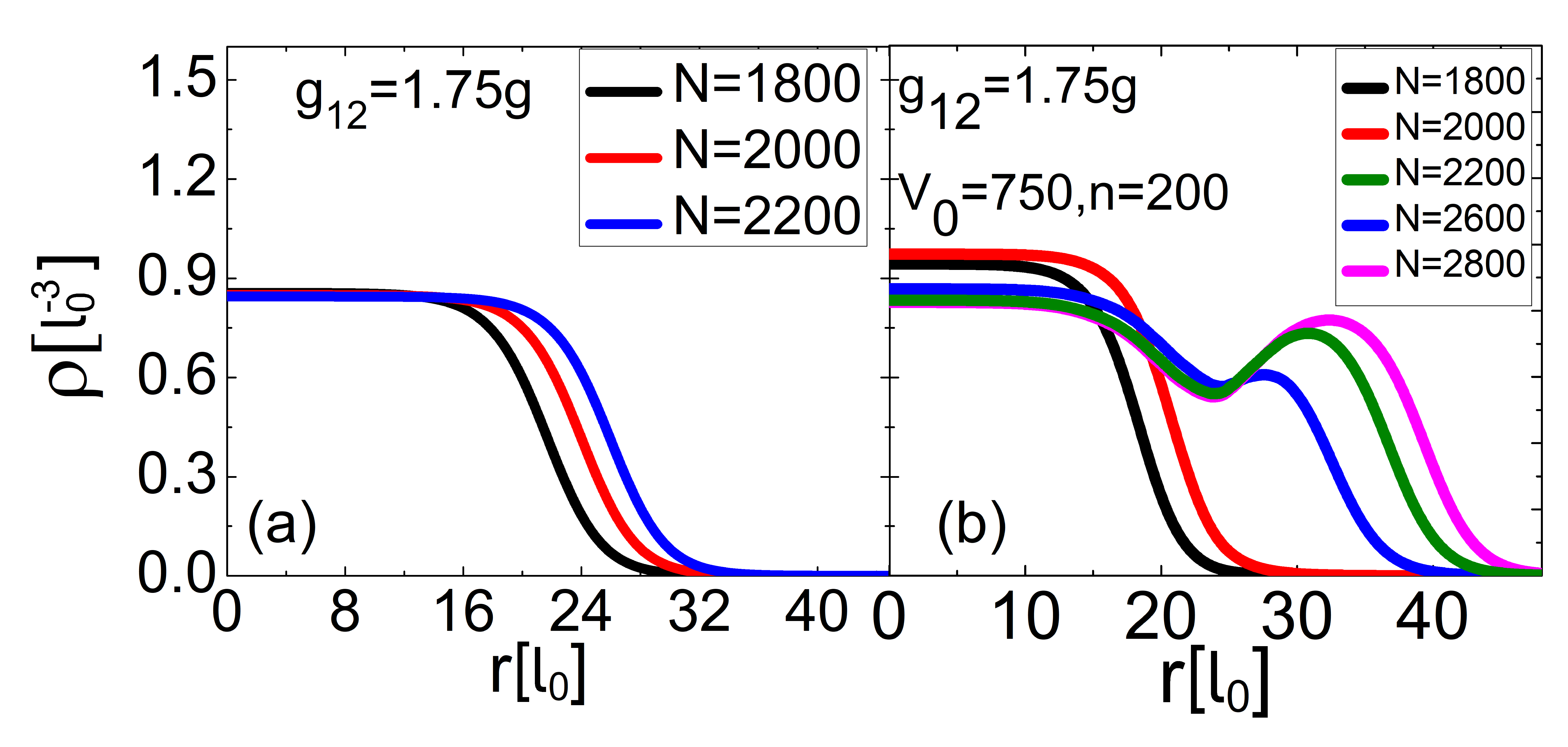}
     \includegraphics[width=0.5\textwidth]  {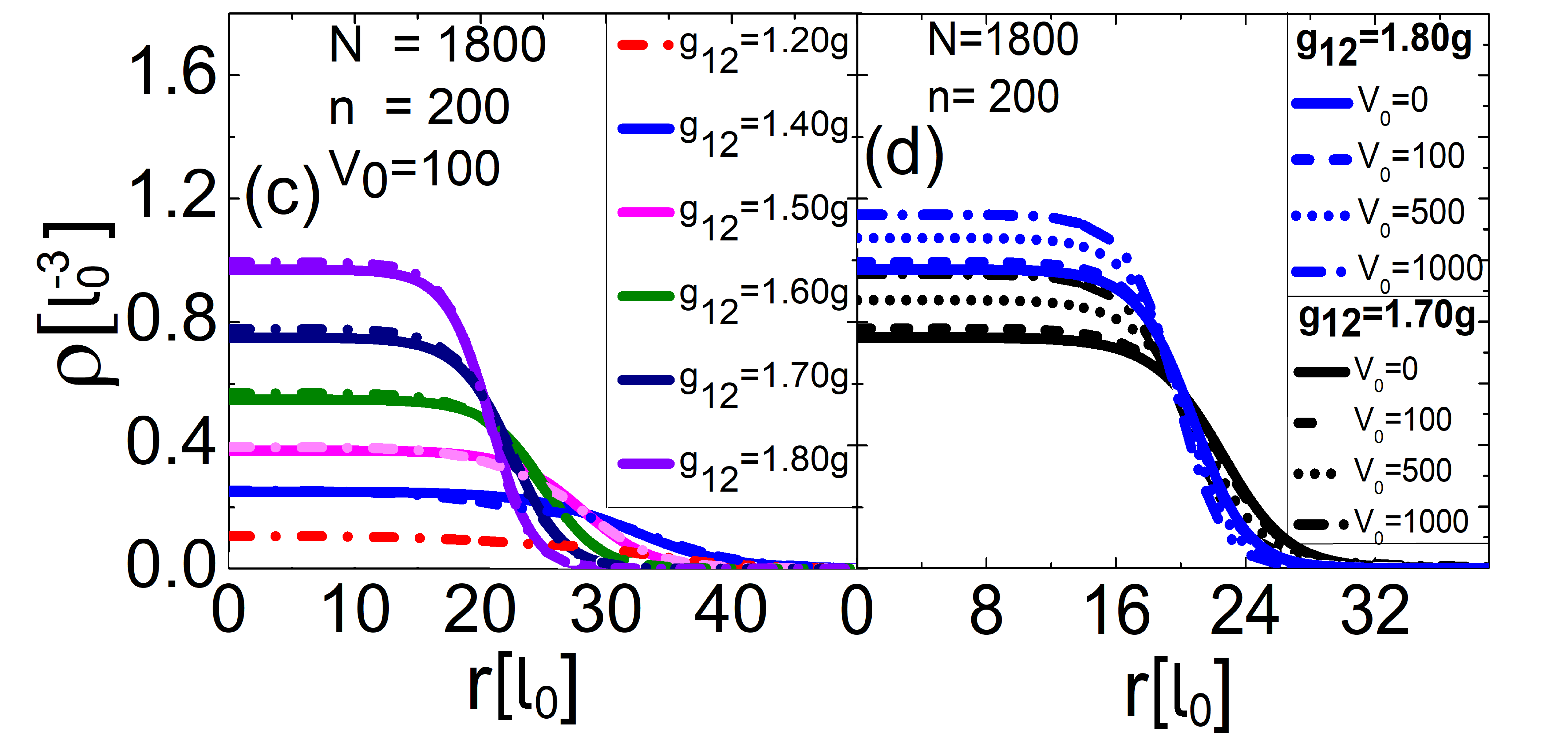}
    \caption{Density profile of the mixture of two condensates($g=1$).
    \textbf{ (a)}:droplet without  Rydberg atom. The size of the droplet increases with the number of particles, keeping the density fixed.
    \textbf{(b)}: density of the doplet is not fixed; it slightly varies with the number of particles. Some atoms will move out of the Rydberg atom if we increase the number of particles beyond a certain limit.
   \textbf{(c)}: we have compared the density of free droplet and liquid phase inside Rydberg atom for various interspecies interactions. The dashed line represents mixtures within the Rydberg atom($N=800$ for $g_{12}=1.20g$).
  \textbf{ (d)}: same as the adjacent picture for different values of electron-atom interaction.}

   \end{center}
   \end{figure}

 The normalization condition is given by $4\pi\int (|\phi_{1}|^{2}+|\phi_{2}|^{2}) d\textbf{r }= N$, where $N$ denotes the total number of particle. 
 Here also, we have applied the semi-implicit CN method to solve the coupled non-linear equation. Again we have considered a sufficiently large distance as the boundary. The calculated density profile is shown in Fig. 3.
 
 
 
 
 
 \subsection{Collective excitation}

Again our condensate is about 40$l_0$ in diameter, whereas the coherence length is about $l_0$, so the size of the condensate is sufficiently large to study the collective excitation of the system in the central region, where the density of the condensate is almost uniform. The excited state wave function is taken as a perturbation over the ground state, $\psi_j^{\mbox{exc}} = \psi_j+\delta \psi_j$. Where \cite{dm_coll},

\begin{footnotesize}
\begin{equation}
  \delta \psi_j = e^{-i\mu t} \left (U_j e^{i\vec{q}\cdot  \vec{r} -i\omega t} + V_j^* e^{-i\vec{q}\cdot  \vec{r} +i\omega t}\right )
\end{equation}
\end{footnotesize}
Where chemical potential is given by \cite{ref10,adhikary},
\begin{eqnarray}
  \mu =4\pi\int_0^\infty \biggl [  \biggr. & \frac{1}{2} \left (|\frac{\partial   \phi_1}{\partial  r}|^2 + |\frac{\partial   \phi_2}{\partial  r} |^2 \right )+\frac{g}{r^2}\left (|\phi_1|^4+|\phi_2|^4 \right )  \nonumber \\
  &\;\;\; -\frac{2g_{12}}{r^2}|\phi_1|^2|\phi_2|^2  + \frac{g_{LHY}}{r^3}(|\phi_1|^5+|\phi_2|^5) \nonumber \\
  & \biggl . + V_0 |\Psi (r)|^{2}(|\phi_1|^2+|\phi_2|^2 ) \biggr ]    dr
\end{eqnarray} 
 where  $\{U_j, V_j\}$ are amplitude of the excitation. If we put these in our GP equations, we will find the equations for $\{U_j, V_j\}$, in the first order approximation of $\{\delta\psi_i\}$ \cite{dm_coll} .

 \begin{footnotesize}
   \begin{eqnarray*}
\left(
    \begin{tabular}{c c c c}
      $H_1+C$ & $-g_{12}\psi_1 \psi_2^*$ & $A \psi_1^2$&$ -g_{12}\psi_1 \psi_2$\\
       $-g_{12} \psi_1^* \psi_2$ & $H_2+C$ & $-g_{12} \psi_1 \psi_2$ &  $B \psi_2^2$\\
       $-A\psi_1^{*2}$ & $g_{12}\psi_1^* \psi_2^*$ &$-H_1-C$ & $g_{12}\psi_1^* \psi_2$\\
       $g_{12} \psi_1^* \psi_2^*$ & $-B \psi_2^{*2}$ & $g_{12} \psi_1 \psi_2^*$ & $-H_2-C$\\
    \end{tabular}
\right)
\left(
\begin{tabular}{c}
$U_1$ \\$U_2$\\$V_1$\\$V_2$ 
\end{tabular}
\right) = \omega \left(
\begin{tabular}{c}
$U_1$ \\$U_2$\\$V_1$\\$V_2$ 
\end{tabular}
\right) 
\end{eqnarray*}

  \end{footnotesize}
where 
\begin{eqnarray}
H_1 &=&  \frac{q^2} {2}+ 2g_{11} |\psi_1|^2- \mu- g_{12} |\psi_2|^2+\frac{5}{2}g_{LHY}|\psi_1|^3 \nonumber \\
H_2 &=&  \frac{q^2} {2}+ 2g_{22}|\psi_2|^2-\mu- g_{21} |\psi_1|^2+\frac{5}{2}g_{LHY}|\psi_2|^3 \nonumber \\
A &=& g_{11}+\frac{3}{2}g_{LHY}|\psi_1|  \nonumber \\
B &=& g_{22}+\frac{3}{2}g_{LHY}|\psi_2|  \nonumber \\
C &=&   V_0 |\Psi|^2 \nonumber
\end{eqnarray}

We obtain the system’s excitation spectra by solving the aforementioned matrix. The following figure displays the excitation spectra in Fig. 4 and Fig. 5\cite{dm_coll,ch_bose} .

     \begin{figure} 
   \begin{center}
        \includegraphics[width=0.5\textwidth]{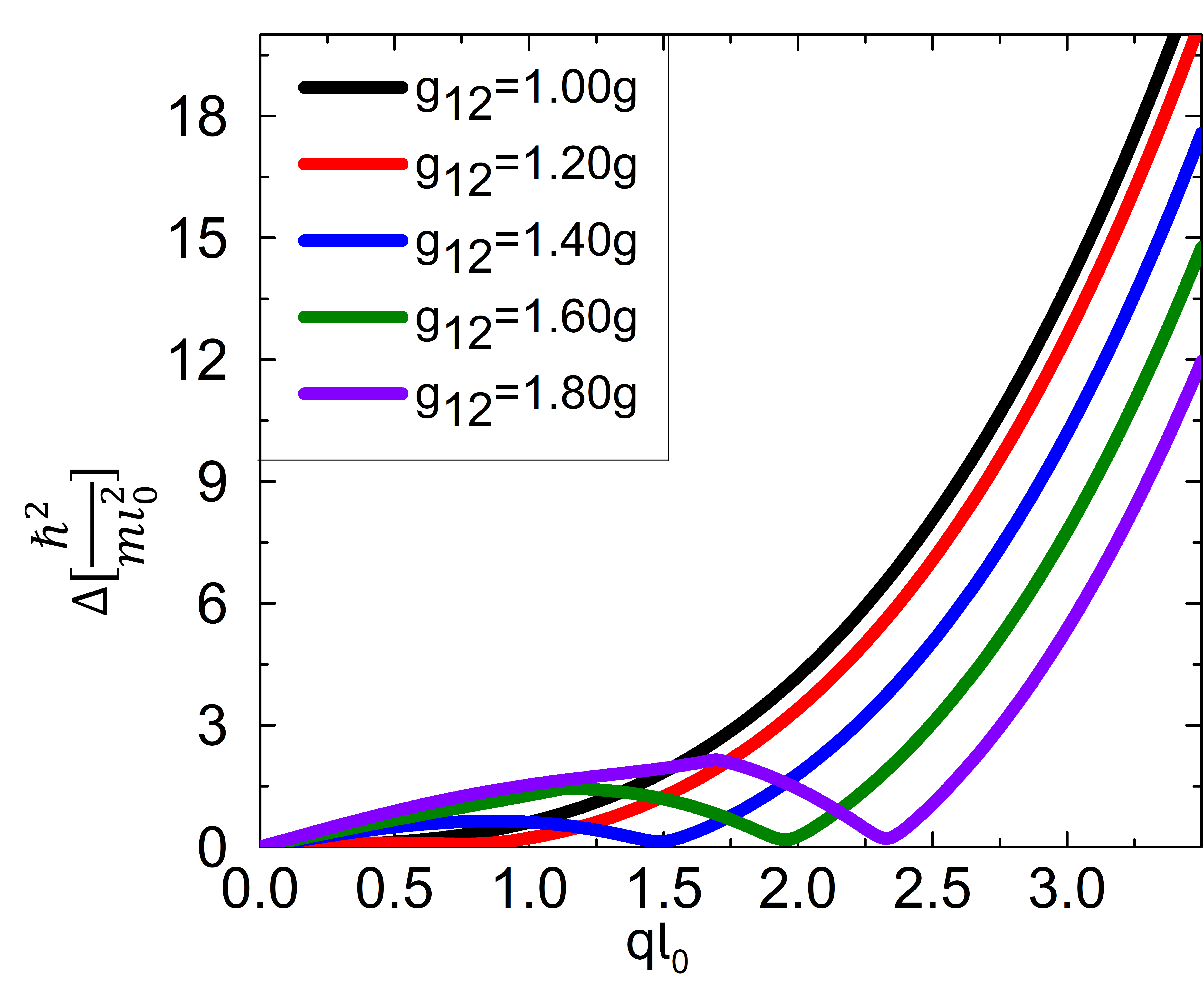}
        
    \caption{Collective excitation of  free droplet for different values of $g_{12}$ and $g=1$. For $g_{12}=1$ (black curve), there is only phonon mode present, and for larger $g_{12}$ there is a mixture of phonon and roton modes.}
       
        \end{center}
  \end{figure} 
 \begin{figure} 
   \begin{center}
\includegraphics[width=0.51\textwidth]{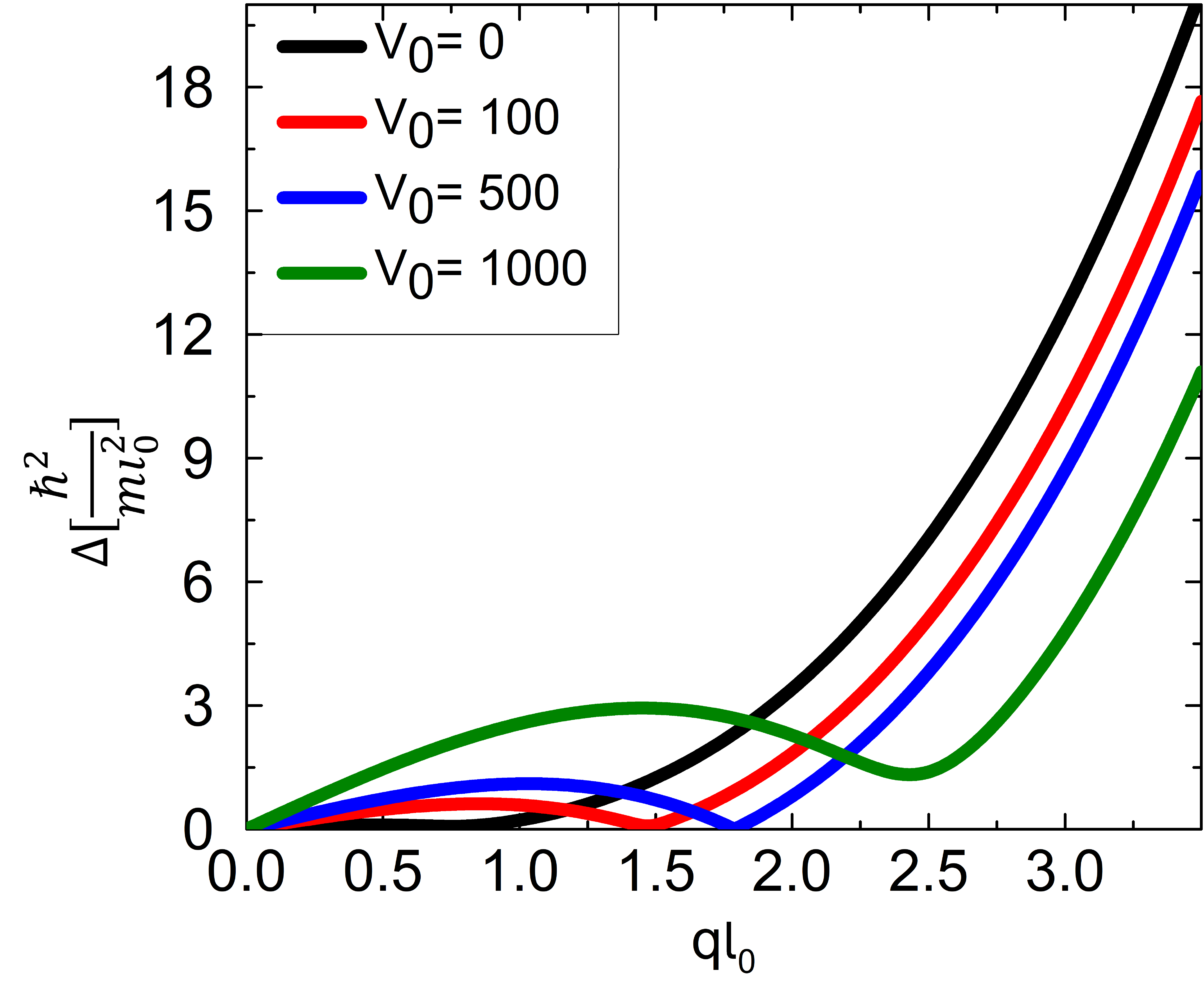}
    \caption{Collective excitation spectra of the quantum liquid in free droplet form (black curve) and confined inside the Rydberg atom for different values of $V_0$,  where $g_{12}=1.2$ and $g=1$. For larger $V_0$, interaction becomes comparatively strong and long-ranged. There is a mixture of phonon and roton modes for larger values of $V_{0}$.}
        
  \end{center}
  \end{figure}

\section*{{\normalsize II.Results and discussions}} 

 \textbf{One species BEC:}\\
The density profile of single component BEC confined in a Rydberg atom has been shown in Fig. 1( For various configurations of the interaction parameter and the number of particles). In Fig. 1(c), we have seen that the Rydberg electron with the principal quantum number of 200 and $V_0 = 4000$ can confine 400 particles, but if the radius of the Rydberg electron orbital is reduced to the principal quantum number of 180, some of the 400 atoms move out of the Rydberg atom. A large fraction of the condensate moves out of the Rydberg atom for a smaller size ($n = 150$). The amount of condensate (number of condensed atoms) confined by the Rydberg electron depends on the interaction between condensed atoms ( Fig. 1(b) ) . We have observed that the Rydberg electron with the principle quantum number 200 can confine up to 1000 atoms with the interaction parameters $g=1$ and $V_0=4000$ (Fig. 1(d) ).

We used perturbations such as equations (4) and (11) since we are considering a sufficiently large system, and we are studying the collective excitation in the central zone, where the density is uniform\cite{dm_coll}.
Collective excitation of weakly interacting uniform BEC of bulk systems is roton less. The collective excitation of the condensate confined by a Rydberg electron is exactly similar to that of a bulked system due to the uniform distribution of the condensate. So the condensate confined in a Rydberg atom may be a suitable finite system to study the properties of bulk condensate (Fig. 2).

\vspace{0.4cm}
\textbf{Two species BEC:}\\
We have solved the coupled GP equation (10) to get the ground state density profile for two species of BECs confined in the Rydberg atom shown in  Fig. (3) for various configurations of interaction and the different number of particles. In our calculation, we have considered $N_1=N_2$ , and the number of particles in the two condensates is equal. We have plotted the total density (individual density is half of the total density). The density of the condensate without Rydberg dressing in the liquid phase is independent of the particle number of atoms (Fig. 3(a)) \cite{qd3}. Where the density rises slightly as a result of the Rydberg electron interaction (Fig. 3(d)) and depends on the number of particles present in the system, the electron can confine a finite number of particles.

In Figure 4, we have studied the effect of $g_{12}$ on collective excitation. For $g_{12}=1$ , excitation spectrum is roton-less. A mixture of roton and phonon modes are developed for larger values of $ g_{12}$ . The roton energy increases and shifts to slightly higher momentum with more attractive interspecies interaction. The collective excitation of two species of condensate confined by a Rydberg electron is shown in Fig. 5. Here we have considered repulsive interaction between the atoms of the same species. The roton mode is developed as the interaction strength $V_0$ increased. For larger values of $V_0$ there is a mixture of phonon and roton modes. The roton shifts to slightly higher momentum with  larger $V_0$.

\end{document}